# Dispersion Relation of Longitudinal Waves in Liquid He-4 in the Framework of Quantum Macroscopic Equations derived from Bohm's Potential

Vincenzo Molinari, Domiziano Mostacci[*]

*Laboratorio di Montecuccolino, Bologna University*
*Via dei Colli,16, I-40136 Bologna (ITALY)*

## Abstract

He-4 is known to become superfluid at very low temperatures. This effect is now generally accepted to be connected with BEC (Bose-Einstein Condensation). The dispersion relation of pressure waves in superfluid He-4 has been determined at 1.1 °K by Yarnell et al., and exhibits a non monotonic behavior - with a maximum and a minimum - usually explained in terms of excitations called rotons, introduced by Landau. In the present work an attempt is made to describe the phenomenon within the bohmian interpretation of QM. To this end, the effects of the intermolecular potential, taken to be essentially of the Lennard-Jones type modified to account for molecule finiteness, are included as a Vlasov-type self-consistent field. A dispersion relation is found, that is in quite good agreement with Yarnell's curve.

**Keywords**: Bohm potential; Intermolecular potential; Dispersion relation; Superfluid He-4; First sound.

**PACS**: 05.30.Jp; 05.20.Dd; 03.75.Kk; 67.25.dt

## 1. Introduction

He-4 is known to exhibit a peculiar behavior at very low temperatures, referred to as superfluidity. At the temperature of 2.18 K liquid helium undergoes a sharp transition, well seen experimentally: for instance, in specific-heat experimental measurements – there the shape of the $(C_V, T)$ curve resembles the Greek letter $\lambda$ and so the transition temperature is often referred to as $\lambda$ point – and in superfluidity. This effect is believed to be connected with BEC (Bose-Einstein Condensation). Standard perfect boson gas calculations, if applied to liquid $He^4$ ($m = 6.65 \cdot 10^{-24}$ g and $n = 2.2 \cdot 10^{22}$ cm$^{-3}$), yield a value for $T_c$ of $3.13$ °K,

[*] correspondinc author: domiziano.mostacci@unibo.it, +39-051-2087719, fax +39-051-2087747

larger than the temperature $T_\lambda$ at the λ-point; however, as was pointed out already by London [1], and is discussed even very recently by Balibar [2], intermolecular forces are bound to change the value of $T_c$ and it is now generally accepted that the transition at the λ-point is really due to BEC.

The anomalous physical behavior of liquid helium has been modeled with a phenomenological theory invoking a two-fluid concept: He-I, the normal liquid helium, and He-II, appearing at the transition temperature. This model was put forth by Tisza [3] and discussed further by London [1] in a seminal paper. Landau [4] developed in another form a theory based on this model, and investigated the problem of the propagation of ordinary sound in liquid helium, in which the two fluids move together to create a pressure wave moving at ca. $240\,m/s$. Landau introduced *ad hoc* characteristic excitations he called *rotons*, and proposed an energy-momentum spectrum of the elementary excitations in liquid helium at temperatures below the λ point that was later substantially confirmed experimentally by Yarnell et al [5], who determined the dispersion relation of sound waves (the so-called *first sound*) in superfluid He-4 at 1.1 °K: the dispersion relation shows a non monotonic behavior - with a maximum and a minimum - usually explained in terms of Landau's rotons. Much work has been done on the roton theory: recent comprehensive overviews can be found, e.g., in [2, 6], see also [7] for a historical recount. In some authors' view, rotons are thought to be phonons of a wavelength close to interatomic spacing; others propose quantized vortices as a model for rotons. In the present work a different approach is taken, and an attempt is made to describe the phenomenon in the bohmian interpretation of QM. To this end, the effect of the intermolecular potential needs to be taken into account: here this potential is taken to be of the Lennard-Jones type with a modification importing a distance of minimum approach to account for the finite size of the molecules – what is proposed is essentially a hybrid between Lennard-Jones and Sutherland potentials. Starting from this potential, a Vlasov-type self-consistent field is then calculated and used in conjunction with a set of quantum macroscopic equations to determine a dispersion relation for longitudinal waves.

## 2. Methods

*2.1 Quantum Macroscopic Equations*

According to De Broglie, Bohm and others, quantum mechanics may be interpreted causally, the wave function playing the role of trajectory generating functions. In contrast to the usual interpretation, Bohm's alternative view leads to individual systems obeying deterministic laws [8]. In Bohm's interpretation, a quantum mechanical potential is introduced

$$U(\mathbf{r},t) = -\frac{\hbar^2}{2m}\frac{\nabla^2 R}{R} \qquad (1)$$

where R(r,t) is a non-negative real function, the modulus of the wave function $\Psi$

$$\Psi(\mathbf{r},t) = R(\mathbf{r},t)\exp\left\{\frac{iS(\mathbf{r},t)}{\hbar}\right\} \qquad (2)$$

This introduces a force $\mathbf{F}_Q$ generated by the quantum Bohm potential which, recalling that $R^2 = n$ with n the number density, can be rewritten as

$$\mathbf{F}_Q = -\frac{\partial}{\partial \mathbf{r}}U_B = \frac{\partial}{\partial \mathbf{r}}\left(\frac{\hbar^2}{2m}\frac{\nabla^2\sqrt{n}}{\sqrt{n}}\right) \qquad (3)$$

Quantum kinetic equations (QKE) can be written in this framework [11], and from those QKE the following quantum macroscopic equations (QME) were derived in a previous work by the present authors [12]

$$\frac{\partial n}{\partial t} + \frac{\partial}{\partial \mathbf{r}}\cdot(n\mathbf{w}) = 0 \qquad (4a)$$

$$\frac{\partial \mathbf{v}_0}{\partial t} + \mathbf{w}\cdot\frac{\partial \mathbf{w}}{\partial \mathbf{r}} + \frac{1}{nm}\frac{\partial}{\partial \mathbf{r}}\cdot\overset{\Rightarrow}{\Psi} - \frac{\mathbf{F}+\mathbf{F}_Q}{m} = 0 \qquad (4b)$$

where **w** is the average velocity, n the number density, m the particle mass, **F** the overall force - other than the quantum force $\mathbf{F}_Q$ - acting at point **r**, and $\overset{\Rightarrow}{\Psi}$ the kinetic pressure tensor

$$\overset{\Rightarrow}{\Psi} = \int m(\mathbf{v}-\mathbf{w})(\mathbf{v}-\mathbf{w})f d\mathbf{v} \qquad (5)$$

It is worth noting that the above two moment equations, (4-a,b), are identical to the first two quantum hydrodynamic equations obtained by Gardner [13] from a moment expansion of the Wigner-Boltzmann equation and used there to investigate semiconductor devices.

The only force (other than $\mathbf{F}_Q$) considered in the present work is the self-consistent field produced by molecular interactions, which in the following will be referred to as $\mathbf{F}_L$: this force is discussed further in the next subsection. The kinetic pressure tensor will be approximated as $p\vec{U}$ with p the scalar pressure and $\vec{U}$ the unit tensor.

Therefore, the moment equations will be written henceforth as

$$\frac{\partial n}{\partial t} + \frac{\partial}{\partial \mathbf{r}} \cdot (n\mathbf{w}) = 0 \tag{6a}$$

$$\frac{\partial \mathbf{w}}{\partial t} + \mathbf{w} \cdot \frac{\partial \mathbf{w}}{\partial \mathbf{r}} + \frac{1}{nm}\frac{\partial p}{\partial \mathbf{r}} - \frac{\mathbf{F}_L + \mathbf{F}_Q}{m} = 0 \tag{6b}$$

Continuing the parallel with Gardner's equations [13], in that case the self-consistent field was that due to the electrostatic force.

*2.2. The self consistent field*

The global effect of molecular interactions is accounted through the Vlasov self-consistent field $\mathbf{F}_L$, which is the cumulative effect, at the point considered, of the forces from all the surrounding particles, weighted over the density distribution of the latter:

$$\mathbf{F}_L(\mathbf{r}_1) = \int_{\mathfrak{R}_3} n(\mathbf{r}_2) \frac{\partial \varphi_{1,2}(\mathbf{r}_1, \mathbf{r}_2)}{\partial \mathbf{r}_1} d\mathbf{r}_2 \tag{7}$$

where $\varphi_{1,2}(\mathbf{r}_1,\mathbf{r}_2)$ is the pairwise interaction potential between molecules located at $\mathbf{r}_1$ and $\mathbf{r}_2$ respectively, and $n(\mathbf{r}_2)$ is the number density at $\mathbf{r}_2$. The Vlasov approach is well suited to investigating wave propagation in a system where the self-consistent field is dominant over pair correlation, so that the double distribution function for a pair of molecules located at $\mathbf{r}_1$ and $\mathbf{r}_2$ can be simplified as the product of the respective single particle distribution functions:

$$f_{1,2}(\mathbf{r}_1,\mathbf{r}_2) \cong f_1(\mathbf{r}_1) \times f_1(\mathbf{r}_2) \tag{8}$$

This is appropriate to the present case as in liquids every molecule interacts simultaneously with all the surrounding ones, and the global effect is preeminent over one-to-one interactions.

In this work, molecules of finite size are considered, rather than point molecules, and specifically hard spheres of diameter $\sigma$: therefore, the usual Lennard-Jones potential needs some correction to account for this finiteness. To this end an approach akin to that of the

Sutherland potential is taken, which combined with a Lennard-Jones model yields the following modified Lennard-Jones potential (henceforth referred to as mLJ)

$$\varphi_{1,2}(r) = \begin{cases} \infty & \text{for } r \leq \sigma \leq r_0 \\ 4\varepsilon\left[\left(\dfrac{r_0}{r}\right)^{12} - \left(\dfrac{r_0}{r}\right)^6\right] & \text{for } r > \sigma \end{cases} \qquad (9)$$

where $r = |\mathbf{r}_1 - \mathbf{r}_2|$, $\varepsilon$ and $r_0$ are the usual parameters of the Lennard-Jones potential and $\sigma$ is the effective diameter of the hard-sphere molecules – a parameter to be adjusted from experimental data - and constitutes a hard-core distance of closest approach of the molecules, measured as distance between molecule centers - see figure 1. The infinite repulsive potential accounts for this distance of closest approach of the molecules.

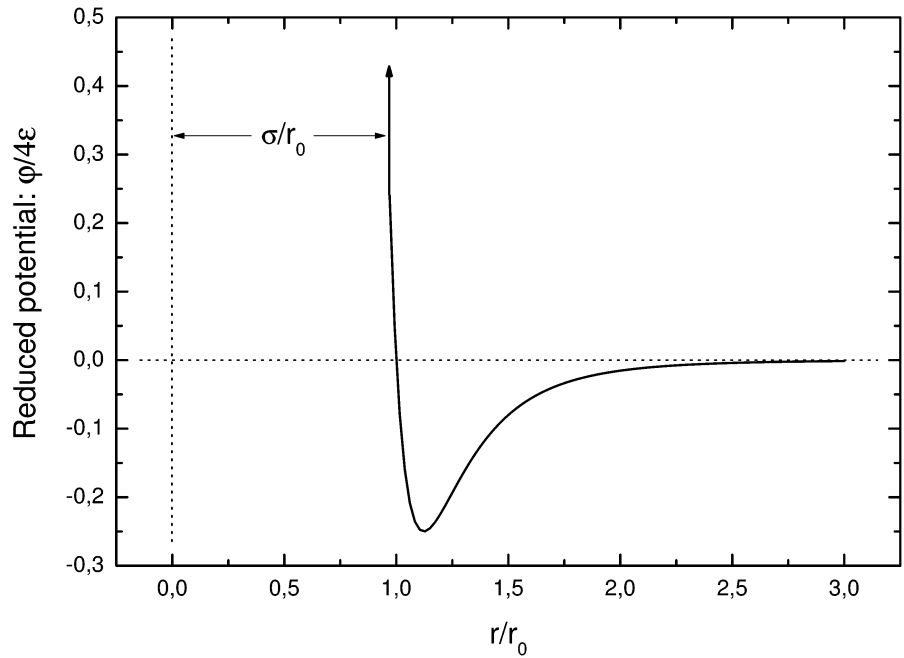

*Figure 1: behavior of the modified L-J potential with intermolecular distance*

This potential, combining the usual L-J potential with the Sutherland potential approach, is discussed at greater length in the Appendix to which the interested reader is referred for further details. There, the self-consistent field is calculated for the one-dimensional case that will be considered here, as

$$F_L(z) \cong \Lambda \frac{dn(z)}{dz} \qquad (10)$$

(to first order) where the parameter $\Lambda$ is given by

$$\Lambda = \frac{16\pi\varepsilon r_0^6}{3\sigma^3}\left[1 - \frac{1}{3}\left(\frac{r_0}{\sigma}\right)^6\right] \qquad (11)$$

*2.3. Wave Propagation and Dispersion Relation*

Consider now a perturbation from an initial, equilibrium state $n = n_0$ and $\mathbf{w} = 0$:

$$n(z,t) = n_0 + \eta(z,t) \qquad \mathbf{w}(z,t) = w(z,t)\hat{\mathbf{z}} \qquad (12)$$

Then, neglecting second order terms, and taking into account the one-dimensional nature of the problem, the first two macroscopic equations become [12]

$$\frac{\partial \eta}{\partial t} + n_0 \frac{\partial w}{\partial z} = 0 \qquad (13a)$$

$$\frac{\partial w}{\partial t} + \frac{1}{n_0 m}\frac{\partial p}{\partial z} - \frac{F_L + F_Q}{m} = 0 \qquad (13b)$$

and, calling $a^2$ the square of the isentropic sound speed, given by

$$a^2 = \frac{1}{m}\frac{\partial p}{\partial n} = \frac{1}{m}\frac{\partial p}{\partial \eta} \qquad (14)$$

Eq. (13b) can be rewritten as

$$\frac{\partial w}{\partial t} + \frac{a^2}{n_0}\frac{\partial \eta}{\partial z} - \frac{F_L + F_Q}{m} = 0 \qquad (15)$$

combining Eqs. (13a) and (15) to eliminate the velocity, an equation for $\eta$ alone is obtained

$$\frac{\partial^2 \eta}{\partial t^2} - a^2 \frac{\partial^2 \eta}{\partial z^2} + \frac{n_0}{m}\frac{\partial}{\partial z}[F_L + F_Q] = 0 \qquad (16)$$

For the present geometry, the quantum force Eq. (3) is

$$F_Q = \frac{\hbar^2}{2m}\frac{\partial}{\partial z}\left[\frac{1}{\sqrt{n}}\frac{\partial^2 \sqrt{n}}{\partial z^2}\right] \qquad (17)$$

and retaining only first order terms the final expression for the Bohm quantum force remains

$$F_Q = \frac{\hbar^2}{4mn_0}\frac{\partial^3 \eta}{\partial z^3} \qquad (18)$$

Finally, the wave equation becomes

$$\frac{\partial^2 \eta}{\partial t^2} - a^2 \frac{\partial^2 \eta}{\partial z^2} - \frac{n_0}{m}\Lambda \frac{\partial^2 \eta}{\partial z^2} + \frac{\hbar^2}{4m^2}\frac{\partial^4 \eta}{\partial z^4} = 0 \qquad (19)$$

Taking both a Fourier transform $z \to k$ and a Laplace transform $t \to s$

$$N(k,s) = \frac{s\tilde{\eta}(k,0) + \tilde{\eta}_t(k,0)}{s^2 + a^2 k^2 - \frac{n_0}{m}\Lambda k^2 + \frac{\hbar^2}{4m^2}k^4} \qquad (20)$$

The dispersion relation is obtained, as usual, equating to zero the denominator of Eq. (20) and setting $s = i\omega$

$$\omega^2 = a^2 k^2 - \frac{n_0}{m}\Lambda k^2 + \frac{\hbar^2}{4m^2}k^4 \qquad (21)$$

## 3. Results and discussion

Wave propagation in liquid He-4 at temperatures below the critical temperature for BEC (Bose-Einstein condensation) will be considered next.

In the quantum dispersion relation Eq. (21) there appears to be no distinction left between fermions and bosons. In the kinetic equation, one prominent source of difference lies in the collision term: the Ueling and Uhlenbeck term differs significantly for the two species, leading, respectively, to the Fermi-Dirac and Bose-Einstein equilibrium distribution functions; however in the present Vlasov self-consistent field approach there is no collision term, so this source of difference is not present. Notwithstanding the lack of a difference in the form of the equation, an important difference is buried in the isentropic sound speed *a*: in the boson case, of interest here, pressure is given by [14]

$$p = \frac{K_B T}{\lambda_{th}^3} g_{5/2}(\alpha) \qquad (22)$$

where

$$\lambda_{th} = \frac{h}{\sqrt{2\pi m K_B T}} \qquad (23)$$

is the mean thermal wavelength, $\alpha$ the fugacity of the gas, related to the chemical potential $\mu$ by

$$\alpha = \exp\left\{\frac{\mu}{K_B T}\right\} \qquad (24)$$

and the Riemann function $g_r$ is given by

$$g_r(x) = \sum_{l=1}^{\infty} \frac{x^l}{l^r} \qquad (25)$$

Below the critical temperature $T_c$ the chemical potential for bosons vanishes identically, i.e., $\alpha \equiv 1$, and the pressure p for $T \leq T_c$ becomes simply [14]

$$p = \frac{K_B T}{\lambda_{th}^3} g_{5/2}(\alpha = 1) \qquad (26)$$

This quantity does not depend on density, therefore $\partial p/\partial n = 0$ and $a^2$ vanishes. This happens because variations in number density at a fixed temperature below the critical one concern only the particles in the BEC, the number of particles in the excited states remains a fixed value determined univocally by the temperature: the particle in the BEC have zero momentum and therefore bring no contribution to the pressure, see e.g. [1,14]. For fermions this is never the case and the $k^2$ term connected with the isentropic sound speed is always present; an equation analogous to Eq. (21) was derived for fermions in a prevoius paper by the present authors [12], where the propagation of waves in plasmas was investigated: in that case the interaction term considered was the Coulomb interaction, instead of the molecular interaction of the present paper.

Coming back to the problem at hand, given Eq. (26) the dispersion relation of elementary excitations for helium at temperatures under the critical point becomes

$$\varepsilon = k\frac{\hbar}{K_B}\left(-\frac{n_0}{m}\Lambda + \frac{\hbar^2}{4m^2}k^2\right)^{1/2} \tag{27}$$

where $\varepsilon = \omega\hbar/K_B$

The interest here is to compare this theoretical spectrum with the experimental one, as obtained in liquid helium by Yarnell et al. [5]. In the region between $k = 0.5 \text{ Å}^{-1}$ and $k = 2.5 \text{ Å}^{-1}$ the energy spectrum possesses a *non-monochromatic* character; in particular the spectrum passes through a maximum value at $k_{MAX} = 1.11 \text{ Å}^{-1}$ and then this maximum is followed by a minimum at $k_{min} = 1.92 \text{ Å}^{-1}$, and then rises again.

As mentioned briefly in section 1, in the region around $p_0 = \hbar k_{min}$ the energy spectrum can be represented by Landau's *roton* spectrum

$$\varepsilon(p) = \Delta + \frac{(p-p_0)^2}{2\mu} \tag{28}$$

where $\Delta$ is a constant and $\mu$ an effective mass. Landau derived the expression of $\varepsilon(p)$ empirically, introducing *rotons,* quantized excitations with specific properties.

In the method proposed here there is no need to introduce idealized quasi-particles, and the experimental results can be interpreted in terms of a simple assumption connected with the self-consistent field.

To this aim, let us observe that the Lennard-Jones intermolecular potential model describes in a simple way a very complex physical problem. The adequacy of this model for providing and analytical description of the typical interparticle potential is confirmed by comparison with a large number of experimental values. This agreement permits to derive empirical values of the parameters of the potential. These parameters are dependent on temperature and density: for instance, when the temperature increases, the particles collide harder with each other and then there is a decrease in the effective hard-core radius (here the parameter $\sigma$), see e.g. [15] for an extensive discussion.

During the propagation of longitudinal wave there is a continuous variation of density and this variation may influence the mLJ parameter $r_0/\sigma$. This effect is likely to be more important for short wavelengths. In the model proposed here this effect is modeled with a first order, linear approximation for $r_0/\sigma$ as a function of the wave number k:

$$\frac{r_0}{\sigma} = a - b \cdot k \qquad (29)$$

If the above expression is introduced in the calculation of $\Lambda$ and hence in Eq. (27), and the usual values for helium for the LJ parameters are used, to wit $\varepsilon = 10.22 \cdot K_B$ and $r_0 = 2.55 \cdot 10^{-10}$ m [16], the lower curve of fig. 2 is obtained. As mentioned in §2.2, the actual value of $\sigma$ needs to be adjusted from experimental data, and hence the coefficient a and b in eq. (29) above need to be adjusted: it can be seen that values $a = 1.25$ and $b = 0.036$ reproduce quite accurately the trend of the experimental curves by Yarnell [5], particularly the locations of minima and maxima; furthermore, if a value 1.6 times larger is used for the parameter group $\varepsilon r_0^3$ (e.g., increasing both $\varepsilon$ and $r_0$ by 12.5%), the experimental curve is reproduced almost exactly, see the second, higher curve in figure 2.

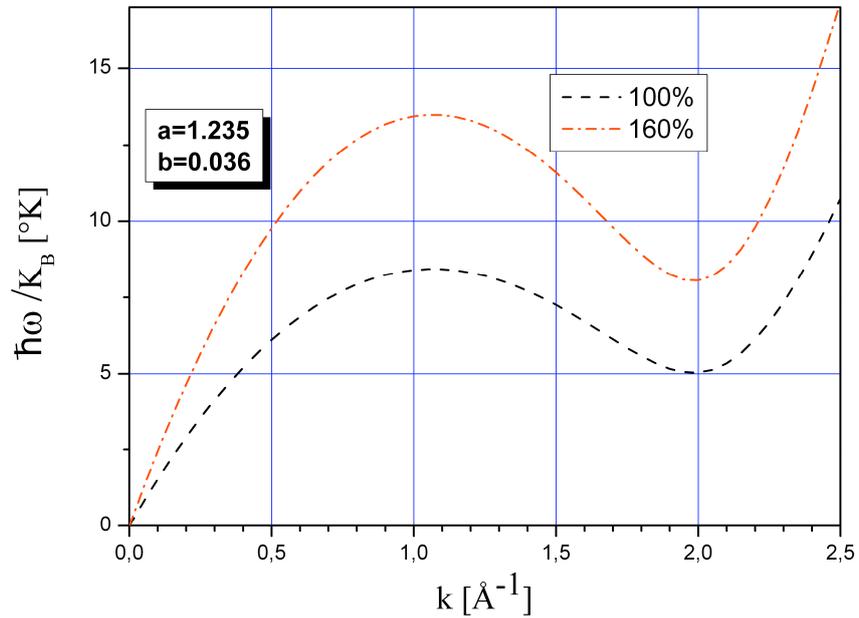

*Figure 2*: *dispersion relation in the present model: the upper curve is obtained by multiplication by 1.6 of the ordinate values of the lower curve*

It should be observed that above the critical temperature the foregoing is no longer true, since the additional $k^2$ term connected to $a^2$ would remain, competing with the self consistent field term and changing the behavior of the dispersion relation

**4. Conclusions**

In the present work an energy-momentum spectrum was obtained for the elementary excitations in liquid helium at temperatures below the $\lambda$ *point*, a spectrum that bears a striking resemblance to what is found experimentally. This result has been reached resorting to the Bohm potential to take into account quantum effects. The global effect of molecular interactions in the liquid state has been accounted for in the usual way, that is, through Vlasov self-consistent field and the modified Lennard-Jones model. Now the question may arise of whether this approach is of more general applicability than just to the problem investigated here: to this end, work is in progress to apply it to studying the so called "second sound" in superfluid helium. This investigation will form the object of a subsequent paper.

**Appendix**

The detailed form of the interaction function can be investigated only through quantum mechanics and much work has been done in this direction [14, 17]. However the problem is very complex and many effects are involved; moreover the structure of the molecules is often not very well known. Therefore the existing results contain significant approximations and are applicable only to specific situations. This being the case, it becomes essential to resort to a phenomenological potential $\varphi_{1,2}$.

In this work, the mLJ intermolecular potential is proposed, as is presented in Eq. 9 in section 2.2 above, and depicted in figure 1.

Now, to calculate the self-consistent force $\mathbf{F}_L$, a field molecule located at the point $(0,0,z)$ will be considered, and the force exerted on this by the whole surrounding liquid will be calculated from the mLJ potential. Again, it is worth stressing that the molecules considered are not point particles, but ibnstead possess a finite size, i.e., a diameter σ, and therefore no one of the other molecules can be located so that its center is closer than a distance σ from the center of the field molecule considered.

Consistently with the 1-dimensional problem posed, a system possessing slab symmetry will be considered, that is one in which density depends only on the z - coordinate. Consider then an elementary volume $dV$ at a location defined by the coordinates $(r,\vartheta,\beta)$ in a spherical reference system centered in the molecule of interest and with the polar axis along the *z* direction, see Fig. 3 [18]

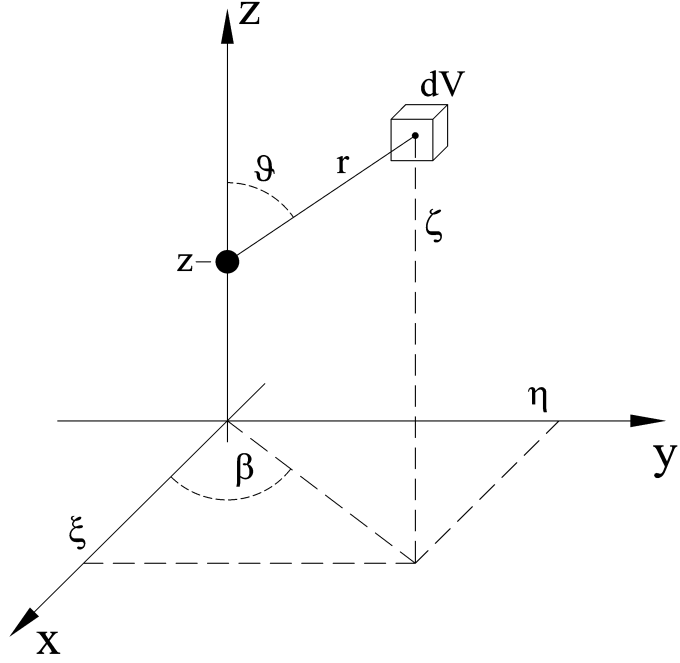

*Figure 3:   Geometry considered for the self-consistent field calculations*

With the geometry in Fig. 3 the force acting on the molecule of interest due to a molecule in $(r,\vartheta,\beta)$ becomes

$$\mathbf{F}_{1,2} = 4\varepsilon\left[\frac{6r_0^6}{r^7} - \frac{12r_0^{12}}{r^{13}}\right]\hat{\mathbf{r}} = \hat{\mathbf{r}}\,4\varepsilon\left[\frac{\alpha r^\alpha}{r^{\alpha+1}}\right]_{\alpha=12}^{\alpha=6} \quad \text{for } r > \sigma \tag{A1}$$

Now, calling $(\xi,\eta,\zeta)$ the Cartesian coordinates of volume dV, the value of $r^{\alpha+1}$ with $\alpha = 12$ or $\alpha = 6$ can be calculated as

$$r^{\alpha+1} = \left[\xi^2 + \eta^2 + (\zeta - z)^2\right]^{\frac{\alpha+1}{2}} \tag{A2}$$

Sines and cosines of the angles in Fig. 3 can be expressed in terms of the Cartesian coordinates $(\xi,\eta,\zeta)$

$$\begin{cases} \sin\vartheta = \dfrac{\sqrt{\xi^2+\eta^2}}{\sqrt{\xi^2+\eta^2+(\zeta-z)^2}} & \sin\beta = \dfrac{\eta}{\sqrt{\xi^2+\eta^2}} \\ \cos\vartheta = \dfrac{\zeta-z}{\sqrt{\xi^2+\eta^2+(\zeta-z)^2}} & \cos\beta = \dfrac{\xi}{\sqrt{\xi^2+\eta^2}} \end{cases} \tag{A3}$$

Consider now the force $d\mathbf{F}_1$ on the field molecule considered, due to the $n(\zeta)d\xi d\eta d\zeta$ molecules contained in the elementary volume dV in figure 3 (Landau's remark [19] on the meaning of "elementary volume" shall be kept in mind, a physical rather than mathematical concept, i.e., small indeed, but large enough to contain a statistically meaningful ensemble): the cartesian component of this force may be rewritten as

$$dF_{1x} = 4\varepsilon\, n(\zeta)d\xi d\eta d\zeta \left[ \frac{\xi r_0^\alpha \alpha}{\left[\xi^2 + \eta^2 + (\zeta-z)^2\right]^{\frac{\alpha}{2}+1}} \right]_{\alpha=12}^{\alpha=6} \tag{A4}$$

$$dF_{1y} = 4\varepsilon\, n(\zeta)d\xi d\eta d\zeta \left[ \frac{\eta r_0^\alpha \alpha}{\left[\xi^2 + \eta^2 + (\zeta-z)^2\right]^{\frac{\alpha}{2}+1}} \right]_{\alpha=12}^{\alpha=6} \tag{A5}$$

$$dF_{1z} = 4\varepsilon\, n(\zeta)d\xi d\eta d\zeta \left[ \frac{(\zeta-z) r_0^\alpha \alpha}{\left[\xi^2 + \eta^2 + (\zeta-z)^2\right]^{\frac{\alpha}{2}+1}} \right]_{\alpha=12}^{\alpha=6} \tag{A6}$$

To obtain the overall force on the reference molecule, integration over the whole volume (in the present setting over $\Re^3$) is to be taken: this needs to be properly treated as the integrand has a pole in $(0,0,z)$. It proves convenient to break the domain up into three non-overlapping volumes: $V_1 = \{\zeta \geq z+\sigma\}$, $V_2 = \{z-\sigma < \zeta < z+\sigma\}$ and $V_3 = \{\zeta \leq z-\sigma\}$. Since $\Re^3 = V_1 \cup V_2 \cup V_3$ and the volumes are non-overlapping the integral over $\Re^3$ equals the sum of the integrals over the 3 volumes $V_1$, $V_2$ and $V_3$. Consider first the integral over $V_1$ of $dF_{1x}$:

$$\int_{z+\sigma}^{+\infty} d\zeta \int_{-\infty}^{+\infty} d\eta \int_{-\infty}^{+\infty} F_{1x} d\xi = 4\varepsilon \int_{z+\sigma}^{+\infty} d\zeta \int_{-\infty}^{+\infty} d\eta \int_{-\infty}^{+\infty} n(\zeta) \left[ \frac{\xi r_0^\alpha \alpha}{\left[\xi^2 + \eta^2 + (\zeta-z)^2\right]^{\frac{\alpha}{2}+1}} \right]_{\alpha=12}^{\alpha=6} d\xi = 0 \tag{A7}$$

It vanishes since the integrand is an odd function of $\xi$. By the same token, the integral of $dF_{1y}$ vanishes being an odd function of $\eta$. Likewise happens in volume $V_3$.

As for the $dF_{1z}$, after some algebra it is found that in $V_1$

$$\int_{z+\sigma}^{+\infty} d\zeta \int_{-\infty}^{+\infty} d\eta \int_{-\infty}^{+\infty} F_{1x} d\xi = 8\pi\varepsilon \left[ r_0^\alpha \int_{z+\sigma}^{+\infty} \frac{n(\zeta)}{(\zeta-z)^{\alpha-1}} d\zeta \right]_{\alpha=12}^{\alpha=6} \tag{A8}$$

and a similar expression in $V_3$, so that the integral over the two volumes $V_1$ and $V_3$ sums to

$$\int_{V_1 \cup V_3} dF_{1z} = 8\pi\varepsilon \left[ -r_0^\alpha \int_{-\infty}^{z-\sigma} \frac{n(\zeta)}{(z-\zeta)^{\alpha-1}} d\zeta + r_0^\alpha \int_{z+\sigma}^{+\infty} \frac{n(\zeta)}{(\zeta-z)^{\alpha-1}} d\zeta \right]_{\alpha=12}^{\alpha=6} \tag{A9}$$

Coming to volume $V_2$, some considerations of a physical, rather than mathematical nature are appropriate: in the first place, no surrounding molecule can come closer than $\sigma$ to the point $(0,0,z)$ where the field particle under consideration is located, due to the finite size of the molecules, so $\sqrt{\xi^2 + \eta^2 + (\zeta-z)^2} \geq \sigma$ always, so it never vanishes; outside of this volume, symmetry considerations dictate that forces from all points of volume $V_2$ balance each other completely, with a vanishing net result.

Putting all the above results together, it is seen that the $\hat{\mathbf{x}}$ and $\hat{\mathbf{y}}$ components of the force vanish, and the only remaining component, along $\hat{\mathbf{z}}$, coincides with the total force, which is thus given by

$$\mathbf{F}_L = \hat{\mathbf{z}}\, 8\pi\varepsilon \left[ -r_0^\alpha \int_{-\infty}^{z-\sigma} \frac{n(\zeta)}{(z-\zeta)^{\alpha-1}} d\zeta + r_0^\alpha \int_{z+\sigma}^{+\infty} \frac{n(\zeta)}{(\zeta-z)^{\alpha-1}} d\zeta \right]_{\alpha=12}^{\alpha=6} \tag{A10}$$

If the density variation is mild, $n(z)$ can be expanded in Taylor series retaining only the first few terms

$$n(\zeta) = n(z) + \frac{dn(z)}{dz}(\zeta - z) + \frac{d^2 n(z)}{dz^2}\frac{(\zeta - z)^2}{2} + \frac{d^3 n(z)}{dz^3}\frac{(\zeta - z)^3}{3!} + \frac{d^3 n(z)}{dz^3}\frac{(\zeta - z)^4}{4!} + O\left[(\zeta - z)^5\right]$$
(A11)

Neglecting terms of order 5 and higher, and substituting into Eq. (16), after some algebra the following equation is obtained:

$$F_L(z) \cong \Lambda_1 \frac{dn(z)}{dz} + \Lambda_3 \frac{d^3 n(z)}{dz^3} \tag{A12}$$

where the coefficients are given by

$$\Lambda_1 = \frac{16\pi\varepsilon\, r_0^6}{3\sigma^3}\left[1 - \frac{1}{3}\left(\frac{r_0}{\sigma}\right)^6\right] \qquad \Lambda_3 = \frac{8\pi\varepsilon\, r_0^6}{3\sigma}\left[1 - \frac{1}{7}\left(\frac{r_0}{\sigma}\right)^6\right] \tag{A13}$$

In what follows only the first term in Eq. (A12), i.e., $\Lambda_1$, will be retained, i.e.,

$$F_L(z) \cong \frac{16\pi\varepsilon r_0^6}{3\sigma^3}\left[1-\frac{1}{3}\left(\frac{r_0}{\sigma}\right)^6\right]\frac{d\,n(z)}{dz} \qquad (A14)$$